\documentclass[12pt]{article}

\usepackage[dvips]{graphicx}
\usepackage{floatflt}
\usepackage{citesort}

\newcommand{\ruim}{\vspace*{5mm}}

\newcommand{\cgo}{CuGeO$_3$}
\newcommand{\cm}{cm$^{-1}$}

\begin{document}

\section*{OPTICAL SPECTROSCOPY ON THE SPIN-PEIERLS COMPOUND \cgo}

\noindent
{PAUL H.M. VAN LOOSDRECHT}
\ruim

\noindent
{\it II Physikalisches Institut der RWTH-Aachen, Templergraben 55, 
            D-52056  Aachen, Germany.}
\ruim

\noindent
\underline{\bf Keywords}: 
Optical Spectroscopy, Magnetic excitations, Spin-Peierls transition.

\noindent
\begin{center}
  \begin{minipage}{12cm}
      {\bf Abstract}\ruim

        An overview is given of Raman and infrared spectroscopic studies of 
        the inorganic spin-Peierls compound CuGeO$_3$, with an emphasis on 
        the magnetic fluctuations in the uniform, dimerized, and high field 
        phases of this quasi one dimensional magneto-elastic compound.
  \end{minipage}
\end{center}

\section{Introduction}

One dimensional quantum spin systems show a variety of interesting physical 
phenomena such as low energy quantum fluctuations, 
pronounced soliton or spinon continua \cite{MUL81}, and the occurrence of 
gapful states in Haldane \cite{HAL83}, spin-Peierls \cite{PYT74}, and 
ladder systems \cite{DAG96}
systems. The magneto-elastic spin-Peierls (SP) transition in organic compounds 
has been studied widely in the 80's \cite{BRA83}. The discovery 
of an inorganic SP compound, \cgo, in 1993 \cite{HAS93} has led to a renewed 
interest in these compounds, which has been boosted once more by the 
discovery of a second inorganic SP compound NaV$_2$O$_5$\ \cite{ISO96}. 
The inorganic nature of these compounds has opened the door to new 
investigations into the nature of SP compounds, which were difficult to 
perform in the past \cite{BOU96}.

The SP transition is the magnetic analogon of the Peierls transition 
in one dimensional metallic systems \cite{PEI55}. It results from 
the instability of 1D antiferromagnetic Heisenberg spin chains at low 
temperatures towards dimerization due to the degeneracy of the ground state 
with the lowest spin excitations at $k_d=\pi$ß \cite{MUL81}. 
The coupling of the 1D magnetic chain to the 3D phonon system may lead to 
a phase transition (the SP transition \cite{PYT74}) which comprises 
a simultaneous dimerization of both the lattice and the magnetic system. 
In the presence of an external magnetic field, the spin system shows 
the tendency to build up a magnetization. As a consequence, the degeneracy 
of the spin excitations with the ground state tends to move away from 
$k_d=\pi$\ \cite{MUL81}, as will the wave vector of the lattice distortion. 
For low fields, however, the distortion is ``pinned'' at 
$k_d=\pi$ due to the presence of {\it umklapp} processes. 
At high enough fields the system will undergo a phase transition to 
a modulated phase by an abrupt jump of $k_d$ away from $\pi$. 
In \cgo\ this has indeed been observed experimentally \cite{KIR95}
In contrast to the dimerized phase, the lattice and spin 
systems now become modulated with a wave vector which is generally 
incommensurate with the periodicity of the undistorted or uniform phase.

Inelastic light scattering (ILS) from magnetic excitations in antiferromagnetic compounds 
has been known for a long time, and 
provides a sensitive tool to study properties such as magnetic 
excitation spectra,  exchange interactions, symmetry, 
and phase transitions \cite{COT86}. 

One of the most important inelastic light scattering mechanisms in 
antiferromagnetic compounds is the so called exchange interaction or 
two magnon scattering process \cite{MOR67,FLE68}. 
The Raman operator for such processes may be written as 
$R=\sum A_{i\vec\delta} (\vec E_{in} \cdot\vec\delta)
(\vec E_{out}\cdot\vec\delta)$\ \cite{ELL69}, 
where $\vec E$ is the optical electric field vector, $i$ labels the 
magnetic spins, and $\vec\delta$ are the connecting vectors to the neighbors of spin $i$. The 
elements $A_{i\vec\delta}$ are symmetry dependent. For $A_g$ symmetries they are 
proportional to the exchange interaction between spin $i$ 
and spin $i+\delta$ in the direction of the polarization the optical electric field. 
In this case the Raman intensity will therefore roughly be proportional 
to the square of the exchange interaction, and in a one dimensional systems one expects 
magnetic scattering only when the incident and scattered light are both polarized along 
the chain direction.

Two magnon processes are also possible in infrared absorption experiments. So far, however,
there have been no reports on the observation of this type of scattering in \cgo. It appears 
that in \cgo\ the most important absorption processes are, like in EPR experiments, due to magnetic dipole 
transitions \cite{BRI94,LOO96B}.

In this contribution I will review some results of inelastic light scattering studies on \cgo\ 
which have been reported over the past five years. The review is inevitably 
somewhat biased by our own work, although I do believe that all key papers have been 
included in this review.

\section{\cgo}

\begin{floatingfigure}[r]{7.0cm}
\includegraphics[width=6.5cm]{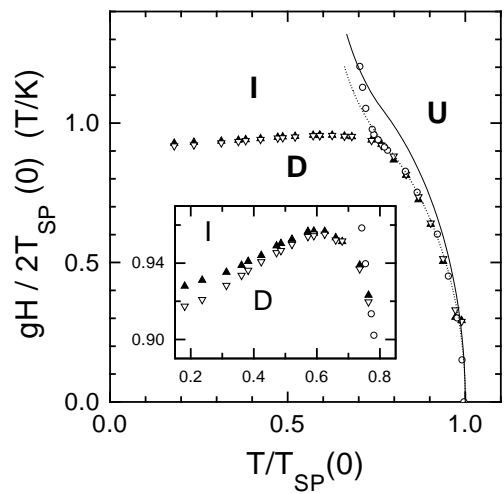}
\caption{
\label{phasediagram}
\it B-T Phase diagram of \cgo\ taken from ref. \cite{LOR97}.}
\end{floatingfigure}
\noindent
A convenient method to produce good quality single crystals of \cgo\ is the 
traveling floating zone method\cite{REV93}. Using this technique one 
easily obtains large (10 $\times$ 5 $\times$ 100 mm$^3$), 
high quality single crystals suitable for use in 
inelastic optical or neutron scattering experiments.

The room temperature structure of \cgo\ is or\-tho\-rhom\-bic with 
space group Pbmm, and unit cell parameters $a=4.81$\  \AA; $b=8.43$\ \AA;
$c=2.95$\  \AA\ \cite{VOL67}. The structure is build of chains of 
planar CuO$_4$ squares running along the $c-$direction, which are separated from each other 
by chains of GeO$_4$ tetrahedra.
The $k=0$ vibrational modes can be classified according to \cite{DEV94,POP95A}:
\begin{eqnarray*}
\Gamma_{vib}&=&4A_g\oplus4B_{1g}\oplus3B_{2g}\oplus B_{3g}\\
&&2A_u\oplus4B_{1u}\oplus6B_{2u}\oplus 6B_{3u},
\end{eqnarray*}
where the {\it gerade} modes are Raman active and the {\it ungerade} $B$ modes 
are infrared active. For a more thorough discussion on the phonon scattering in \cgo\ 
see for instance Popovi\'{c} {\it et al.} \cite{POP95A}. 

The magnetic ions in \cgo\ are the Cu$^{2+}$ ions (spin $S=1/2$) located 
on a fully symmetric site of the structure. 
The exchange interaction in the chain ($c$) direction is 
determined by an almost 90$^o$  Cu-O-Cu superexchange path, leading to a relatively small effective 
nearest neighbor (nn) exchange integral of $J_{c}\sim 120 K$\ \cite{HAS93,NIS94}.  
Frustrating antiferromagnetic next nearest neighbor (nnn) interactions have been used to 
explain the observed deviations of the observed temperature dependence of the magnetic
susceptibility
\cgo\ \cite{CAS95,RIE95}, and may explain the observed temperature dependence of the 
magnetic susceptibility. The values reported for the frustration $\alpha=J_c^{nnn}/J_c^{nn}$ 
are quite high and range from 0.24 to 0.37, with $J^{nn}_c\approx 160$\ K 
\cite{CAS95,RIE95,BUC96}. It has been shown that under 
hydrostatic pressure the frustration may become even larger \cite{BUC96,LOO97}.

The magnetic interactions between the magnetic chains is much smaller, 
though not negligible: $J_b\sim 0.1J_c$; $J_a\sim -0.01J_c$. The presence
of the interchain interaction leads to a sizable dispersion 
of the spin-excitations along the $b-$direction, and consequently to the occurrence of 
two magnetic gaps in the spin-Peierls phase \cite{NIS94,REG96,BRI94,LOO96B},
situated at different points in the Brillouin zone \cite{UHR97} 
($\Delta_{SP}=17$\ \cm\ at $k=(k_b,K_c)=(0,0)$, $(\pi,\pi)$, and $\Delta_b=44$ \cm\ 
at $k=(0,pi)$, $(\pi,0)$).

The phase diagram of \cgo\ has been determined by various authors \cite{HAS93A,LOR97,LOR98}.
It more or less follows the generic phase diagram predicted for a spin-Peierls compound \cite{BUL78,CRO79A}, 
although there are some serious deviations, in particular at high magnetic fields \cite{LOR98}. 
The phase diagram after \cite{LOR97} is reproduced in figure \ref{phasediagram}. It comprises 
a dimerized phase below 14 K and 12.5 T, an incommensurate phase below 9 K and above 12 T, 
and a uniform phase for higher temperatures. The uniform phase itself can be divided into two 
regimes \cite{LOO96}. At low temperatures ($T<T_{max}$, where 
$T_{max}\approx60$\ K $\approx J_c/2$\ is the temperature where the 
magnetic susceptibility reaches its maximum.) there exist short range correlations in the spin system, 
whereas at high temperatures there are only weak correlations.

\section{The uniform phase}

\begin{floatingfigure}{8.5cm}
\includegraphics[width=8.0cm]{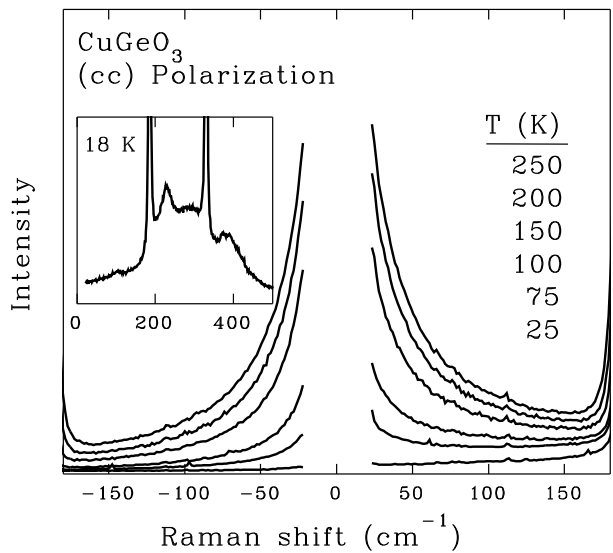}
\caption{
\label{quasi}
\it (cc)-Polarized Stokes and anti-Stokes 
Raman spectra of \cgo\ in the uniform phase showing a broad, strongly temperature dependent 
quasi-elastic scattering. The inset shows the scattering continuum in the 
short range order regime of the uniform phase in $(cc)$-polarization.}
\end{floatingfigure}

The 12 Raman and 15 infrared active pho\-nons of the high temperature phase
have all been observed experimentally \cite{DEV94,KUR94,POP95A}.
In addition, the acoustic modes have been studied by Brillouin 
spec\-tro\-sco\-py \cite{YAM95}. Spin-depen\-dent scattering in the HT-regime
has been observed only in $(cc)$-polarized Raman scattering spectra \cite{LOO96,KUR97} 
as a strong quasi-elastic peak (see figure \ref{quasi}). This polarization selection 
rule is in excellent agreement with the one dimensional character of \cgo, and continues to hold 
in the dimerized and high field phases as well. 
The origin of the quasi elastic scattering is at present not very well understood. 
Initially it has 
been pro\-pos\-ed \cite{LOO96} that this scattering may be due one dimensional 
diffusive behavior of the spin-exci\-tations since there is no short or long range order at high 
temperature. Later it has also been suggested that the scattering is due to magnetic 
energy relaxation \cite{KUR97}. This latter type of scattering should have 
a distinct dependence on the momentum transfer in the scattering process. Experimentally,
however, the quasi-elastic scattering has no momentum transfer dependence at all \cite{MIC98}.
Another possibility may be that the quasi-elastic scattering is due to spin-phonon processes, 
which have not yet been considered in literature.

Upon decreasing temperature (but still $T>T_{sp}$) the intensity of the QE peak 
decreases strongly. At low temperatures ($T_{sp}<T<J/2$) the QE peak has 
almost vanished, and one now observes only a broad band of excitations peaking 
around 230 \cm\  (see inset figure \ref{quasi}) \cite{LOO96,LEM96}. The development of this continuum is due to the 
increasing short range spin-spin correlations in the (quasi-)one-dimensional spin system, 
which in turn leads to the formation of the spin-wave or spinon continuum \cite{MUL81}. 
The reason for the spin-scattering in the $A_g(cc)$ channel is not so clear. For a one 
dimensional nn Heisenberg chain one would not expect any scattering at all in this 
channel, since the exchange Raman operator commutes with the Hamiltonian itself. 
Muthukumar {\it et al.} \cite{MUT96} therefore proposed that the scattering is due to the 
presence of frustration in \cgo. Under this assumption they indeed found a remarkable agreement 
between their theory and the experimental results. Other origins of the scattering may be the 
two dimensionality of the system, or the presence of strong magneto-elastic interactions.
Interestingly, it has been shown recently that taking spin-phonon interactions into account
may lead to frustration in the magnetic system under the assumption that the typical phonon 
energies are much larger than the typical spin excitation energies \cite{UHR98}.

\section{The dimerized phase}
Raman spectroscopic studies of the dimerized phase have been reported by several 
groups \cite{KUR94,LOO96,LEM96,LOA96}. The spin-Peierls transition leads to several new 
features in the (cc) polarized Raman spectra (see figure \ref{dim}). 
In the first place, three zone boundary phonons appear in the (cc) spectra of the dimerized phase, 
at 105, 370, and 820 \cm, which are activated either by magneto-elastic interactions 
or by the structural distortion \cite{KUR94,LOO97A,FIS98}. 
Note that the 105 \cm\  mode shows a strong Fano lineshape distortion due 
to interactions with the spin excitations \cite{LOO96}. 
\begin{floatingfigure}{8.5cm}
\includegraphics[width=7.5cm]{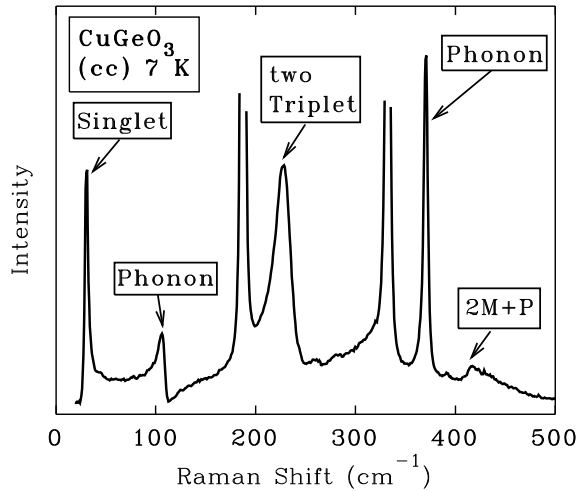}
\caption{
\label{dim}
\it Part of the (cc)-Polarized Raman spectrum of \cgo\ in the dimerized phase.}
\end{floatingfigure}

Secondly, a sharp, somewhat asymmetric mode appears in the Raman spectrum around 
30 \cm. This mode has been assigned to the $k=0$ singlet mode, which is a bound state 
of two antiparallel triplet excitations \cite{UHR96,BOU97}. The binding energy is of the order of 1-2 \cm, 
and increases upon applying hydrostatic pres\-sure \cite{LOO97}. 

Thirdly, a well defined continuum of scattered intensity 
can be observed in the spectra starting around 30 \cm, 
and showing a sharp maximum at 228 \cm. This scattering continuum is generally 
thought to be due to two magnon exchange scattering from the well defined triplet 
excitations in SP phase. Recent inelastic neutron scattering results, however, have 
shown that one of the phonons involved in the SP transition has an energy of 
about 228 \cm\ \cite{BRA97}. Therefore, it seems likely that the peak at 228 \cm\ 
is due to mixed vibrational/magnetic excitations. 

Finally there is a broad structure in the (cc) spectra above 230 \cm, which 
extends up to 500 \cm. This structure is most likely due to spin-phonon scattering processes. 
The peak at 417 \cm\  may, for instance, be assigned to a process where one $A_g$(330 \cm) phonon, 
and two $k=(0,\pi)$ or $(\pi,0)$ gap (44 \cm) excitations are created. 

It appears that one magnon scattering can not be observed for \cgo, probably because of 
the small spin-orbit coupling in Cu. One magnon scattering would appear in the inelastic 
light scattering spectra as a peak around 17 \cm\  ($=\Delta_{SP}$). What has been observed 
at low energies using Brillouin spectroscopy is a thermally activated mode, arising from scattering 
between excited states \cite{ELS97}, as well as an impurity induced mode in doped 
single crystals \cite{ELS98,LOA96}.

The pressure dependent phase diagram of \cgo\ has also been studied \cite{ADA91,JAY95,GON96,LOO97}.
At about 6 GPa there is a structural phase transition to a monoclinic phase \cite{ADA91}, above
which no magnetic excitations have been observed in the Raman spectra \cite{GON96,LOO97}.
The SP transition temperature of \cgo\ strongly increases upon applying hydrostatic 
pressure \cite{WIN95,TAK95,NIS95A}. This has also been observed using Raman spectroscopy 
by monitoring the 370 \cm\  mode \cite{GON96} or the singlet excitation \cite{LOO97}. 
The decrease of the peak energy of the two magnon continuum in the SP phase upon 
increasing pressure has led to the conclusion that the magnetic frustration in 
\cgo\ has a large positive pressure coefficient, and that the SP phase transition 
is largely driven by frustration \cite{LOO97}. This is consistent with earlier 
thermodynamic experiments which led to the same conclusions \cite{BUC96}.

There have been a few reports on infrared spectroscopy in the dimerized 
phase \cite{LOO96B,LI96,DAM97}. 
The main observations are a new absorption 
peak at 44 \cm\ \cite{LOO96B,LI96}, also seen in electron paramagnetic resonance 
spectroscopy \cite{BRI94}, and 
the occurrence of new phonon modes \cite{DAM97} in the SP phase. Also the temperature 
dependence of the phonon spectrum has been studied \cite{LI96,DAM97}.

The 44 \cm\  absorption has been assigned to a magnetic dipole transition 
from the singlet ground state to the triplet gap mode. 
This interpretation, however, is problematic since the $k=0$ 
singlet-triplet gap is about 17 \cm\ \cite{NIS95,UHR97}, and since normally 
singlet-triplet transitions are magnetic dipole forbidden. 
Nevertheless, the absorption behaves as a triplet in
the sense that it indeed splits in a magnetic field (see figure \ref{ir}) \cite{LOO96B,LI96}, and its energy 
corresponds closely to the singlet-triplet gap at $k=(0,\pi)$ or $(\pi,0)$\ \cite{NIS94,REG96}. The exact 
mechanism of the absorption process is at present unclear.

\section{The high field phase}
\begin{floatingfigure}{8.5cm}
\includegraphics[width=7.5cm]{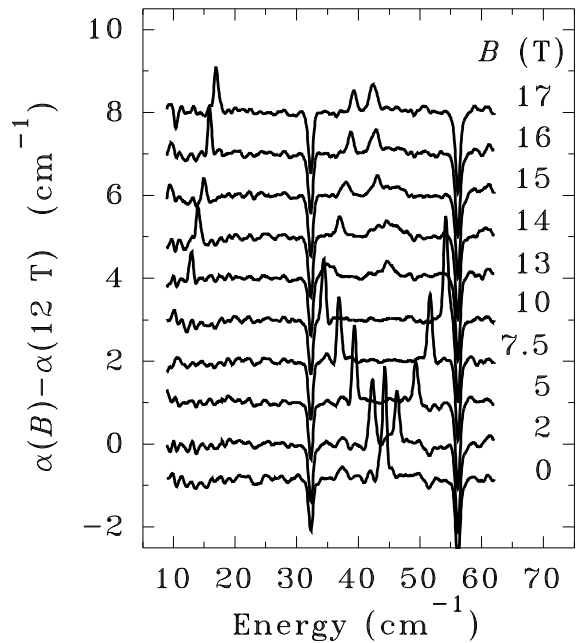}
\caption{
\label{ir}
\it Absorbance difference spectra $\alpha(B)-\alpha(12 {\rm \ T})$\ at $T=2.3$\ K 
in the dimerized and incommensurate phases of \cgo. (ref. \cite{LOO96B})}
\end{floatingfigure}
The high field, or incommensurate phase of \cgo\ occurs for 
magnetic fields above 12.5 T and temperatures below about 
9 K (see figure \ref{phasediagram}). There have been only a few 
reports on optical spectroscopy in the IC phase \cite{MMM95,LOO97A,LOO96B,LI96}.
In the dimerized phase, the presence of well defined triplet excitations 
led to a scattering continuum in the $A_g(cc)$ channel at energies between 
the singlet response at 30 \cm, and the two magnon peak at 228 \cm. 
In the IC phase, the singlet response disappears due to the closure (or 
decrease in energy ?) of the spin-Peierls gap. 
A clear two magnon peak is, though weaker, still observed \cite{MMM95,LOO97A}, 
indicating that there still exist more or less
well defined, propagating triplet excitations in the center of the Brillouin zone 
with energies of the order of $\pi J$. 
Also the SP induced phonons observed in the D phase at 105 and 370 \cm\  persist  
in the IC phase, but again with a much weaker intensity. In fact, the temperature 
dependence of either the two-magnon or the induced phonons can be used to monitor 
the second order U-D or U-IC transitions, as well as the first order D-IC transition.
At this latter transition, the intensities show, as expected for a first order 
transition, a step-like decrease upon entering the IC phase \cite{MMM95}.

Like the disappearance of the singlet response in Raman spectroscopy, also the 
44 \cm\  gap mode observed in infrared spectroscopy \cite{LOO96B,LI96} disappears in 
the IC phase. Instead, one now observes a sharp absorption with an 
low energy scaling linearly with the magnetic field (see figure \ref{ir}, mode 
at 13-17 \cm).
In contrast to the non-magnetic singlet ground state of the D phase, 
one expects a magnetic ground state in the IC phase.
The low energy mode at 13 \cm\  (for $B=13$\ T) has therefore been assigned to 
transitions between levels of the Zeeman splitted ground state. 
Far infrared spectra have also given direct evidence for a modulated structure 
in the IC phase by the appearance of absorption processes involving magnetic 
and vibrational excitations with $k=k_{ic}$ (where $k_{ic}$\ is the modulation vector)
in the 30-50 \cm\ region (see figure \ref{ir}). 
The observation of an additional mode at 45 \cm\ 
for fields just above the D-IC phase transition strongly indicates
the existence of a discommensurations regime in the IC phase for fields 
between 12.5 and 16 T. 

\section{Conclusions}

The overview given in this contribution shows that optical 
spectroscopies can give quite a lot of insight into the properties of 
low dimensional magnetic and/or magneto-elastic compounds. Many things 
have been discussed or touched upon such as spin and energy diffusion, 
spin excitation continua, spin-phonon processes, propagating triplet excitations, 
singlet bound states, magnetic frustration, phase transitions, and 
incommensurate structures. Nevertheless, some things had to be left out. 
The most important omissions are probably the temperature and pressure dependent 
optical studies the phonon spectra, and the investigations 
of the strong influence of that non-magnetic 
substitutions have on the properties of \cgo\ \cite{LEM97,MCG97,DEV97,FIS98,ELS98,DAM97,DAM97A}.

\section*{Acknowledgments}

I thank Prof. J.-P. Boucher for introducing me to the subject of 
spin chain systems and to \cgo, and for many fruitful discussions. 
Sincere thanks are also due to the group of Prof. A. Revcolevschi and Prof. G. Dhalenne 
for growing excellent \cgo\ single crystals, for making them available to us, 
and for stimulating discussions. Finally, thanks are due to all the people at 
the Grenoble High Magnetic Field Laboratory, the II Physikalisches Institut der RWTH-Aachen, 
and at the Institut f\"ur Theoretische Physik and II Physikalisches institut der
Universit\"at zu K\"oln for their numerous contributions and fruitful discussions.

\end{document}